\documentclass{jfm}
\usepackage{graphicx}
\usepackage{epstopdf, epsfig}

\usepackage[bookmarks=false, pdfstartview={FitH}, colorlinks = true, citecolor = blue, linkcolor = blue, urlcolor = blue,hyperfootnotes=true]{hyperref}
\usepackage{cleveref}
\usepackage{bm}

\DeclareFontFamily{U}{euc}{}
\DeclareFontShape{U}{euc}{m}{n}{<-6>eurm5<6-8>eurm7<8->eurm10}{}%
\DeclareSymbolFont{AMSc}{U}{euc}{m}{n} 
\DeclareMathSymbol{\umu}{\mathord}{AMSc}{"16} 

\usepackage{upgreek}

\shorttitle{Transport Barrier with Microorganisms in Chaotic Flows}
\shortauthor{R. Ran and P. E. Arratia}

\title{Enhancing Transport Barriers with Swimming Microorganisms in Chaotic Flows}

\author{Ranjiangshang Ran\aff{1,2} \and Paulo E. Arratia\aff{1}
   \corresp{\email{parratia@seas.upenn.edu}}}

\affiliation{\aff{1} Department of Mechanical Engineering and Applied Mechanics, University of Pennsylvania, Philadelphia, PA 19104, USA
\aff{2} Department of Physics, Emory University, Atlanta, GA 30322, USA}

\begin{document}

\maketitle

\begin{abstract}
We investigate the effects of bacterial activity on the mixing and transport properties of a passive scalar in time-periodic flows in experiments and in a simple model. We focus on the interactions between swimming \textit{E. coli} and the Lagrangian Coherent Structures (LCSs) of the flow, which are computed from experimentally measured velocity fields. Experiments show that such interactions are non-trivial and can lead to transport barriers through which the scalar flux is significantly reduced. Using the Poincar\'e map, we show that these transport barriers coincide with the outermost members of elliptic LCSs known as Lagrangian vortex boundaries. Numerical simulations further show that elliptic LCSs can repel elongated swimmers and lead to swimmer depletion within Lagrangian coherent vortices. A simple mechanism shows that such depletion is due to the preferential alignment of elongated swimmers with the tangents of elliptic LCSs. Our results provide insights into understanding the transport of microorganisms in complex flows with dynamical topological features from a Lagrangian viewpoint.
\end{abstract}


\section{Introduction}
Many microorganisms live in environments characterized by currents (e.g., oceans, rivers, human intestines), and flow mediates many important microbial processes such as infection \citep{Costerton_Science_1999,Stocker_JBac_2016}, uptake of nutrients \citep{Taylor_Science_2012,Musielak_JFM_2009}, and reproduction \citep{Zimmer_J_Exp_Biol_2007,Riffell_PNAS_2011}. Flow exerts forces and torques on microorganisms that affect their motility dynamics and spatial distribution \citep{Guasto_Annu_Rev_2012,Stocker_Annu_Rev_2019}. It also controls the transport of essential molecules including nutrients, oxygen, and signals of mates and predators \citep{Kim_Nat_MicroBio_2016}. Microorganisms, in turn, can adapt their swimming behavior to these physical and chemical gradients \citep{Stocker_PNAS_2008}. Such interactions can lead to non-trivial phenomena such as rheotaxis \citep{Hill_PRL_2007, Marcos_PNAS_2012, Mathijssen_NC_2019}, gyrotaxis \citep{DeLillo_PRL_2014, Borgnino_JFM_2018}, and chemotaxis \citep{Locsei_BMB_2009,Stocker_MMBR_2012}, to name a few examples.


Even simple shear flows, when coupled to cell motility and morphology, can give rise to complex transport and cell motility behavior \citep{Rusconi_NatPhys_2014,Durham_Science_2009,Saintillan_JFM_2015,Gustavsson_PRL_2016}. For instance, fluid shear can cause a torque that can rotate elongated cell body and can result in accumulation of motile bacteria and phytoplankton in high-shear-rate regions of the flow \citep{Rusconi_NatPhys_2014,Stocker_JRSI_2015,Saintillan_JFM_2015}. Near surfaces, shear flows can orient flagellated cells against the flow, causing bacteria and spermatozoa to swim upstream \citep{Tung_PRL_2015, Zaferani_PNAS_2018,Mathijssen_NC_2019}. Away from surfaces, shear gradients can trap bottom-heavy gyrotactic swimmers at certain depth of water column, causing the formation of intense cell assemblages called ``thin layers'' \citep{Durham_Science_2009,Durham_NC_2013,Gustavsson_PRL_2016}. In unsteady and/or complex flows, transport of microorganisms shows intriguing phenomena such as aggregation, dispersion, and pattern formation \citep{Torney_PRL_2007,Khurana_PRL_2011,Khurana_PoF_2012,Zhan_JFM_2014,Qin_Sci_Adv_2022}, but are less understood. In time-periodic flows, simulations show that elongated swimmers can be trapped or repelled by elliptic islands depending on their shape and swimming speed \citep{Torney_PRL_2007}; such trapping effects can lead to a reduction in long-term swimmer transport \citep{Khurana_PRL_2011}. Recent experiments and simulations show that microswimmers can be trapped, repelled, or dispersed by vortices depending on the dimensionless path-length and swimming speed \citep{Qin_Sci_Adv_2022}. In isotropic turbulence, simulations show that elongated swimmers, while remaining rather uniformly distributed, exhibit preferential alignment with instantaneous Eulerian fields such as local velocity \citep{Borgnino_PRL_2019}, vorticity \citep{Zhan_JFM_2014} and velocity gradient \citep{Pujara_JFM_2018}.


Recently, it has been shown that Lagrangian Coherent Structure (LCS) can be a useful concept to understand the transport properties of swimming microorganisms in complex flows in both numerical simulations and experiments \citep{Khurana_PoF_2012,Dehkharghani_PNAS_2019,Ran_PNAS_2021,Si_Fang_PoF_2021,Si_Fang_PRF_2022,Solomon_SwIMs_2022}. Simulations in chaotic flows show that elongated swimmers align with repelling LCSs of hyperbolic fixed points \citep{Khurana_PoF_2012}, while later numerical studies show that elongated active particles have a much stronger alignment with attracting LCSs \citep{Si_Fang_PoF_2021,Si_Fang_PRF_2022}, similar to passive elongated particles \citep{Parsa_PoF_2011}. Experimental investigations that examine the interactions of swimming organisms and flow LCSs are few but show some intriguing phenomena. Experiments in model porous media show that bacteria align and accumulate near attracting LCSs and induce filamentous density patterns \citep{Dehkharghani_PNAS_2019}. In time-periodic flows, experiments show that the accumulation of bacteria near the attracting LCSs can attenuate stretching and hinder large-scale transport, although small-scale mixing is locally enhanced \citep{Ran_PNAS_2021}. Most, if not all, previous studies focus on attracting and repelling LCSs associated with the flow hyperbolic fixed points. That is not surprising since one expects large levels of strain near or around those fixed points. Less understood are swimmer interactions with elliptic LCSs, i.e., vortex-like flow dynamical features \citep{Haller_ARFM_2015,Haller_JFM_2016,Farazmand_PhysD_2016}. That is the focus of this manuscript.

Here, we experimentally investigate the effects of bacterial activity on the mixing and transport properties of a passive scalar in a time-periodic flow in experiments and simulations. We focus on the interaction of swimming bacteria (\textit{Escherichia coli}) with the elliptic LCSs of the flow. Results show that such interaction leads to transport barriers through which the fluxes of the passive tracer are significantly reduced. By constructing the Poincar\'e map from velocimetry data, we show that these transport barriers coincide with the outermost member of elliptic LCSs, or namely, Lagrangian vortex boundaries (LVBs). We further test these results in numerical simulations and find that elliptic LCSs repel elongated swimmers and lead to swimmer accumulation outside (or swimmer depletion inside) the Lagrangian vortices. A simple mechanism shows that the repulsion of swimmers is due to the preferential alignment of elongated swimmers with the tangents of elliptic LCSs. Overall, these results can be useful in understanding the transport of microorganisms in chaotic flows with elliptic dynamical features or nontrivial vortex structures.

\begin{figure}
  \centerline{\includegraphics[width=3.37in]{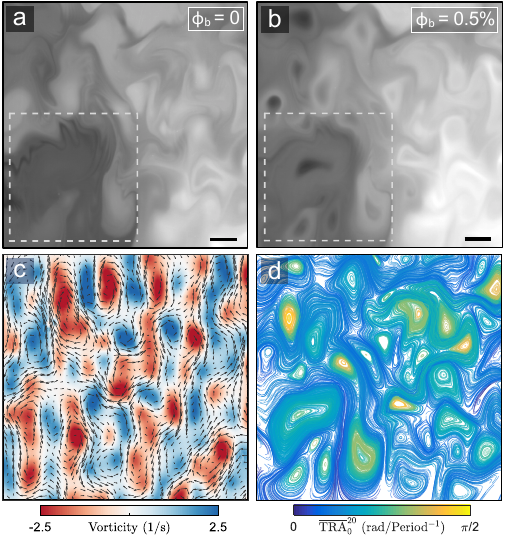}}
  \caption{Photographs of the dye concentration field for chaotic mixing in (\textit{a}) buffer solution ($\phi_b=0$) and (\textit{b}) active suspension ($\phi_b=0.5\%$). Images are taken at $N=300$ periods after the start of the experiments. The imaged region is $60\times60$~mm${}^2$; the scale bar represents 6~mm. The Reynolds number and path-length of the flow are $Re=7.2$ and $p=1.0$, respectively. (\textit{c}) Vorticity (color code) and velocity (arrows) fields of the flow, corresponding the the first peak of a time period. (\textit{d}) Poincar\'e map of the flow in the same imaged region as the dye field. The map is colored by the trajectory rotation average (TRA) calculated over 20 time periods.}
  \label{fig1}
\end{figure}

\section{Methods}
Experiments are performed in the flow cell setup where a 2-mm thin conductive fluid layer is placed above an array of permanent magnets arranged in a disordered pattern \citep{Voth_PRL_2002,Ran_PNAS_2021,Ran_PRF_2022}. As a sinusoidal forcing (electrical current, 0.2~Hz frequency) is imposed on the fluid layer, the magnetic field induces a Lorentz force and creates spatially disordered vortex patterns [see flow field in Fig. \hyperref[fig1]{1(c)}]. The resulting flow is characterized by two parameters: the Reynolds number and the path-length. The Reynolds number is defined as $Re=U\!L/\nu$, where $U=1.2$~mm/s is the average flow speed, $L=6.0$~mm is the characteristic length scale determined by the average spacing of the permanent magnets, and $\nu=1.0$~mm$^2$/s is the fluid kinematic viscosity (water-like). The path-length is the normalized mean displacement of a typical fluid parcel in one forcing period, defined as $p=UT/L$, where $T=5$~s is the forcing period. Here, the Reynolds number and the path-length are $Re\approx7.2$, $p\approx1.0$; these conditions are known to lead to chaotic advection in this system \citep{Ran_PNAS_2021,Voth_PoF_2003}.

Two main types of experiments are performed: dye mixing and particle tracking velocimetry (PTV). Dye mixing experiments are performed by labelling half of the fluid layer with a passive dye tracer ($6.25\times10^{-5}~$M sodium fluorescein). Initially, the labeled and unlabeled portions are separated by a physical barrier. As the flow begins, the barrier is lifted, and the labeled fluid progressively penetrates the unlabeled portion with time. The dye concentration field is recorded by a CMOS camera (Flare 4M180) at 5 frames/s with a resolution of $2000\times2000$ pixels. Particle tracking experiments are performed by seeding the fluid with 100-$\umu$m large fluorescent polystyrene particles; the Stokes number of these fluorescent particles is $\mathcal{O}(10^{-4})$, indicating good tracer fidelity. Particle positions are recorded by the CMOS camera at 30 frames/s with a resolution of $1200\times1200$ pixels. Particle trajectories are obtained using an in-house tracking algorithm \citep{Crocker_1996}; these trajectories are then used to obtain the velocity fields from 6th-order polynomial fitting. Because the flow is time-periodic, we can combine particle positions at a given phase (relative to the forcing) to obtain up to 80,000 precise particle positions at each phase, which yields high spatial resolution (0.002 of the field of view), excellent temporal resolution (0.007 of a flow period), and velocities accurate to a few percent. Two different velocity maps are obtained from separate experiments, one in the presence and the other in the absence of bacteria.

Active suspensions are prepared by adding a strain (wild-type K12 MG1655) of \textit{Escherichia coli} to an aqueous buffer solution of 2\% KCl and 1\% NaCl by weight; the swimming motility of \textit{E. coli} does not seem to be affected by the salts \citep{Ran_PNAS_2021}. The strain of \textit{E. coli} has a swimming speed of 10--20~$\umu$m/s and a rod-shaped body of on average 2~$\umu$m long and 0.5~$\umu$m in diameter. We note that the P\'eclet numbers of the bacteria and the dye are both considerably large, with $Pe_{b}=U\!L/D_b\sim\mathcal{O}(10^6)$, and $Pe_{d}=U\!L/D_d\sim\mathcal{O}(10^4)$, where $D_b$ and $D_d$ are the effective diffusivities of the swimming \textit{E. coli} and the dye, respectively. Bacteria are cultured in Luria–Bertani (LB Lennox, Sigma-Aldrich) liquid media at 37${}^{\circ}$C overnight for 12 to 14 hours to attain a stationary phase of a number density of approximately $10^9$ cells/mL. The stationary-phase culture is centrifuged at 5,000 revolutions per minute for 3.5 minutes and resuspended into the buffer solution to attain a number density of $1.25\times10^{10}$ cells/mL or a bacterial volume fraction of $\phi_b=0.5\%$. This bacterial volume fraction ($\phi_b=0.5\%$) is considered dilute \citep{Kasyap_PoF_2014,Ran_PNAS_2021}, and large scale collective behavior/motion is not expected.

\section{Results and Discussion}
\subsection{Experimental Results}
Figures \hyperref[fig1]{1(a)} and \hyperref[fig1]{1(b)} show sample snapshots of dye mixing in the buffer solution ($\phi_b=0\%$) and active suspension ($\phi_b=0.5\%$), respectively. Both snapshots are taken after $N=300$ periods of forcing, with dye initially confined on the right half of the images at $N=0$. The concentration field reveals complex patterns and underlying vortex structures, which are similar in the buffer and the active suspension. However, a main difference is the existence of regions devoid of dye near the center of the vortex structures in the active suspension [Fig. \hyperref[fig1]{1(b)}], while similar regions are dyed in the buffer case [Fig. \hyperref[fig1]{1(a)}]. The regions devoid of dye near the center of the vortex appear primarily in the left half of the image that is initially not covered with dye. This suggests that the interplay of microbial activity and vortex structures leads to (enhanced) transport barriers through which the dye fails to penetrate or penetrates much more slowly. A video of the dye mixing processes in the buffer and the active suspension are shown in Movie 1.

To further understand the interaction between activity and vortex structures, we plot in Fig. \hyperref[fig1]{1(d)} the flow Poincar\'e map obtained from PTV experiments. Lines in the map are stroboscopic particle trajectories that connect a particle’s initial position to its next position after a period ($T=5~$s). These trajectories are obtained by numerical integration of fluid particles in the experimentally measured velocity fields. The Poincar\'e map reveals several nested families of closed trajectories, whose behaviors mimic Kolmogorov–Arnold–Moser (KAM) tori in Hamiltonian dynamical systems \citep{Ottino1989}. These nested torus families define Lagrangian coherent vortices known as elliptic Lagrangian coherent structures (elliptic LCSs), with each torus acting as a barrier that blocks tracer (dye) transport within the vortex \citep{Katsanoulis_PRF_2020,Haller_JFM_2020,Aksamit_Haller_JFM_2022}. Unlike hyperbolic LCSs, elliptic LCSs are rotation-dominated regions that do not experience substantial (fluid) stretching. We note that Lagrangian coherent vortices are fundamentally different from the vortices defined by Eulerian fields such as vorticity. While we find a striking correspondence between the dye mixing patterns [Figs. \hyperref[fig1]{1(a)} and \hyperref[fig1]{1(b)}] and the Lagrangian coherent vortices on the Poincar\'e map [Fig. \hyperref[fig1]{1(d)}], such similarity is absent for the Eulerian vorticity field in Fig. \hyperref[fig1]{1(c)}.

To quantitatively locate the elliptic LCSs, we calculate the trajectory rotation average (TRA) on the Poincar\'e map, as shown by the color code in Fig. \hyperref[fig1]{1(d)}. The TRA characterizes the average angular velocity of a particle trajectory and is defined as:
\begin{equation}
    \overline{\mathrm{TRA}}_{0}^{N}(\mathbf{x}_0)=\frac{1}{NT} \sum_{i=0}^{N-1} \cos ^{-1} \frac{\langle\dot{\mathbf{x}}_i,\dot{\mathbf{x}}_{i+1}\rangle}{\left|\dot{\mathbf{x}}_i\right|\left|\dot{\mathbf{x}}_{i+1}\right|}.
\end{equation}
Here, $\dot{\mathbf{x}}_i$ is a stroboscopic velocity related to the rate of change of the position of a particle on the Poincar\'e map and $N$ is the number of periods over which the quantity is calculated. Although the TRA was originally defined for actual particle trajectories \citep{Haller_Chaos_TRA_2021}, we used it here for the stroboscopic trajectories in the extended phase space of the Poincar\'e map. We interpolate the values of the TRA onto a uniform spatial grid to obtain the TRA field (see Movie 2). In two-dimensional flows, elliptic LCSs can be located from the closed convex contours of the TRA field \citep{Haller_Chaos_TRA_2021,Aksamit_Haller_JFM_2022}. Since elliptic LCSs are nested families, we identify the outermost convex TRA contours as the boundaries of Lagrangian coherent vortices, and the centroid of the innermost convex TRA contours as elliptic fixed points. For concision, we refer to the outermost members of elliptic LCSs as Lagrangian vortex boundaries (LVBs).


\begin{figure}
  \centerline{\includegraphics[width=3.37in]{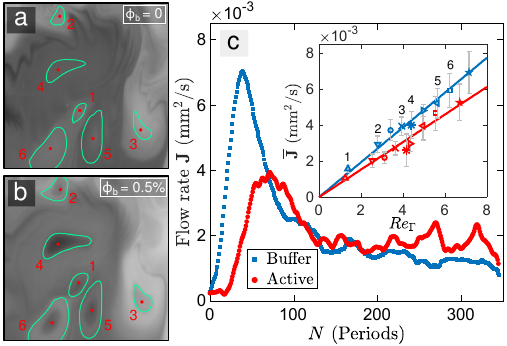}}
  \caption{Enlarged photographs of the dye concentration field for (\textit{a}) buffer and (\textit{b}) active suspension, in the dashed square regions shown in Fig. \hyperref[fig1]{1}. Green contours are Lagrangian vortex boundaries (LVBs) identified from the TRA field and red dots are elliptic fixed points. (\textit{c}) Dye flow rate, $\mathbf{J}$, as a function of time $N$, for the vortex with the label ``4''. Inset: time-averaged dye flow rate, $\overline{\mathbf{J}}$, as a function of the vortex Reynolds number $Re_\Gamma$, for the buffer (blue) and the active suspension (red). Different markers represent different vortices in the flow field. Error bars are standard deviations.}
\label{fig2}
\end{figure}

Figures \hyperref[fig2]{2(a)} and \hyperref[fig2]{2(b)} show the LVBs (green contours) and elliptic fixed point (red dots) identified from the TRA field, superimposed on dye mixing patterns in the buffer and the active suspension, respectively. Although LVBs in chaotic flows are known to be transport barriers for convection \citep{Katsanoulis_PRF_2020,Haller_JFM_2020,Aksamit_Haller_JFM_2022}, the LVBs in the buffer case [Fig. \hyperref[fig2]{2(a)}] contain dye inside due to diffusion. On the other hand, the LVBs in the active case [Fig. \hyperref[fig2]{2(b)}] enclose regions devoid of dye near the elliptic fixed points. This suggests that bacterial activity enhances the strength of LVBs as transport barriers in chaotic flows. We also notice some differences in the size and shape of the LVBs for the buffer and active cases. Since the LVBs are calculated from experimentally measured velocity fields, this indicates that microbial activity also modifies the underlying velocity field.

To quantify the strength of the transport barriers, or the reduction in dye transport, we calculate the dye flow rate into the Lagrangian vortices using the following mass balance equation:
\begin{equation}\label{eqn. flow rate}
    \frac{\mathrm{d}}{\mathrm{d}t}\int\limits_{S(t)}C\,\mathrm{d}A =\oint\limits_{B(t)}D(\mathbf{\nabla} C\cdot\mathbf{n})\,\mathrm{d}l-\oint\limits_{B(t)}C(\mathbf{v}-\mathbf{v}_{B})\cdot\mathbf{n}\,\mathrm{d}l.
\end{equation}
Here, $S(t)$ and $B(t)$ denotes the enclosed area and the contour of the LVBs, $C$ is the dimensionless dye concentration field (normalized by the maximum fluorescence intensity), $D$ is the dye diffusivity, $\mathbf{n}$ is the normal vector of LVBs, $\mathbf{v}$ is fluid velocity, and $\mathbf{v}_{B}$ is the velocity induced by the motion and deformation of LVBs. The first and second terms on the right-hand side of Eq. \ref{eqn. flow rate} are the surface integrals of the dye fluxes through the LVBs due to diffusion and convection, respectively. The left-hand side of Eq. \ref{eqn. flow rate} gives a simple way to calculate dye flow rate as the rate of change of the total dye enclosed by the LVBs. Figure \hyperref[fig2]{2(c)} shows the dye flow rate, $\mathbf{J}$, as a function of time, for vortex with label ``4'' in the buffer and the active suspension. We find a significant reduction of dye flow rate into the LVBs with bacterial activity, especially at earlier times ($N<150$). At later times ($N>200$), the dye flow rate in the active case is larger than that in the buffer, as a result of a larger diffusive flux due to large dye concentration gradients.
 
We now explore the generality of the reduction of dye flow rate in active suspensions. We start by calculating the circulation of a Lagrangian vortex, defined as 
\begin{equation}
    \Gamma(t) = \oint\limits_{B(t)}\mathbf{v}\cdot\,\mathrm{d}\mathbf{l}=\int\limits_{S(t)}\omega_z\,\mathrm{d}A,
\end{equation}
where $\omega_z$ denotes the vorticity component in the out-of-plane direction. We then define the the vortex Reynolds number as: $Re_\Gamma = \left\vert\overline{\Gamma}\right\vert/\nu$, where $\overline{\Gamma}$ denotes the time-averaged circulation over a period, and the absolute value is included since $\overline{\Gamma}$ can be positive or negative depending on the sign of the vorticity. We calculate $Re_\Gamma$ for each Lagrangian vortex in Figs. \hyperref[fig2]{2(a)} and \hyperref[fig2]{2(b)}; the labels correspond to the magnitude of their $Re_\Gamma$. Inset of Fig. \hyperref[fig2]{2(c)} shows the time-averaged dye flow rate, $\overline{\mathbf{J}}$, as a function of the vortex Reynolds number $Re_\Gamma$, for the buffer and active cases. We find that in both cases the average dye flow rates scale linearly with $Re_\Gamma$. This is because both $\overline{\mathbf{J}}$ and $Re_\Gamma$ are linearly proportional to the area of the Lagrangian vortex. More importantly, however, we find that the slope of the scaling is larger in the buffer case compared to the active case. This result shows that the presence of bacteria enhances the barriers for scalar transport into the flow Lagrangian vortices.

\subsection{Numerical Simulations}
To understand how swimming microorganisms interact with an imposed time-periodic flow, we perform numerical simulations of swimming particles using the experimentally measured velocity field. Microorganisms are modeled as axisymmetric ellipsoids with a constant swimming speed $v_s$, in the direction $\mathbf{q}$ along their symmetry axis. The swimmer's position $\mathbf{x}$ is modelled as:
\begin{equation}
    \dot{\mathbf{x}} = \mathbf{v}_f(\mathbf{x},t) +  v_s \mathbf{q},    
\end{equation}
where $\mathbf{v}_f$ is experimentally measured fluid velocity in the active suspension, and $v_s=20~\umu$m/s is the swimming speed of \textit{E. coli}. As a control, simulations of elongated passive particles are also performed with $v_s=0$ in the velocity field measured in the buffer. The swimmer's orientation is described by Jeffery's equation \citep{Jeffery1922}:
\begin{equation}
    \dot{\mathbf{q}}=[\mathbf{W}(\mathbf{x}, t)+\Lambda \mathbf{D}(\mathbf{x}, t)] \mathbf{q}-\Lambda[\mathbf{q} \cdot \mathbf{D}(\mathbf{x}, t) \mathbf{q}] \mathbf{q},
\end{equation}
where $\mathbf{D}$ and $\mathbf{W}$ are the symmetric and skew-symmetric parts of the velocity gradient tensor, $\nabla\mathbf{v}_f$. Here, $\Lambda = (1-\alpha^2)/(1+\alpha^2)$ is a shape factor, with $\alpha$ being the swimmer's aspect ratio. We assume $\alpha=0.25$ for rod-shaped \textit{E. coli}, which leads to $\Lambda\approx0.88$. 

\begin{figure}
  \centerline{\includegraphics[width=3.37in]{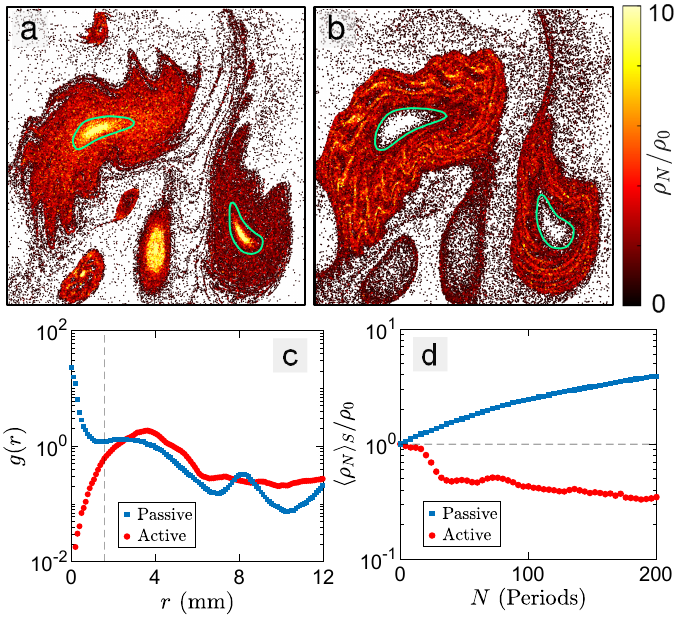}}
  \caption{Spatial distributions of (\textit{a}) passive particles, and (\textit{b}) active particles, at time $N=150$. Particles are colored by their normalized local number density, $\rho_N/\rho_0$. Active particles are depleted from the vortex and accumulate outside the LVBs (green contours), while this depletion is not present for passive particles. (\textit{c}) Radial distribution function $g(r)$ calculated from the elliptic fixed points of the Lagrangian vortex with the label ``4''; the dashed line is the nominal radius of the vortex. (\textit{d}) Spatially averaged normalized number density within the LVBs, $\langle\rho_N\rangle_S/\rho_0$, as a function of time $N$ for the same vortex as in (\textit{c}).}
\label{fig3}
\end{figure}

Initially, both passive and active particles are uniformly distributed in the flow field with random orientations. As simulations begin, passive and active particles begin to develop complex patterns following the morphology of Lagrangian vortices (Movie S3). Figures \hyperref[fig3]{3(a)} and \hyperref[fig3]{3(b)} show the spatial distribution of passive and active particles at $N=150$, respectively. The color code in the plots represents the local particle number density, $\rho_N$, normalized by the initial number density, $\rho_0$. We find that active particles deplete within and aggregate outside the LVBs. This suggest that Lagrangian vortices repel elongated swimmers. By contrast, no depletion is observed for passive particles. Accumulation is not expected for passive particles in a 2D incompressible flow, and aggregation of non-motile bacteria has not been observed in previous experiments \citep{Ran_PNAS_2021}. The accumulation of passive particles seen here is likely due to a small departure from an (ideal) divergence-free experimental velocity field (see Supplementary Materials). The passive particle simulation serves as a control to show that the repulsion of active particles by the Lagrangian vortices is not an intrinsic feature of the experimental velocity field itself. Rather, it is the result of the interaction between Lagrangian vortices and elongated swimmers. 

The repulsion of rod-shaped swimmers by the LVBs is quantified by the radial distribution function: $g(r)=\langle\rho_N (r)\rangle/\rho_0$, where the angle bracket denotes a radial average in all directions calculated from the elliptic fixed points. The function $g(r)$ represents the probability of finding a particle at a radial distance $r$ from the elliptic fixed point of a Lagrangian vortex. Figure \hyperref[fig3]{3(c)} shows $g(r)$ for both passive and active particles in the Lagrangian vortex with the label ``4'' in Fig. \hyperref[fig2]{2}. The dashed line in Fig. \hyperref[fig3]{3(c)} is a nominal radius of Lagrangian vortex, defined as: $r_n=\sqrt{A_S/\pi}$, where $A_S$ is the area enclosed by the LVBs. We find that the likelihood of finding an active particle within the nominal radius of a Lagrangian vortex is smaller than that of a passive particle and vice versa for outside the nominal radius. This result further corroborates the repulsion of elongated active particles by Lagrangian vortices.

To quantify the time evolution of the repulsion of elongated swimmers, we calculate the average number density within the LVBs, $\langle\rho_N\rangle_S$, normalized by the initial number density $\rho_0$. Figure \hyperref[fig3]{3(d)} shows $\langle\rho_N\rangle_S/\rho_0$ as a function of time $N$. Results show that $\langle\rho_N\rangle_S$ decreases nonmonotonically with time for active particles, and drops to only 20\% of the initial number density $\rho_0$ at $N=200$. This suggests that most active particles escape and accumulate outside the Lagrangian vortex as mixing progresses. In the control case, $\langle\rho_N\rangle_S$ for passive particles increases with time, which again suggests the repulsion of elongated particles by LVBs is not an intrinsic feature of the velocity field itself. Although symmetric and circular Eulerian vortices have been observed to repel elongated swimmers \citep{Torney_PRL_2007,Ran_PNAS_2021,Qin_Sci_Adv_2022}, here we extend these results to a more general case of asymmetric and non-circular vortices from Lagrangian criteria.

\begin{figure}
  \centerline{\includegraphics[width=1\textwidth]{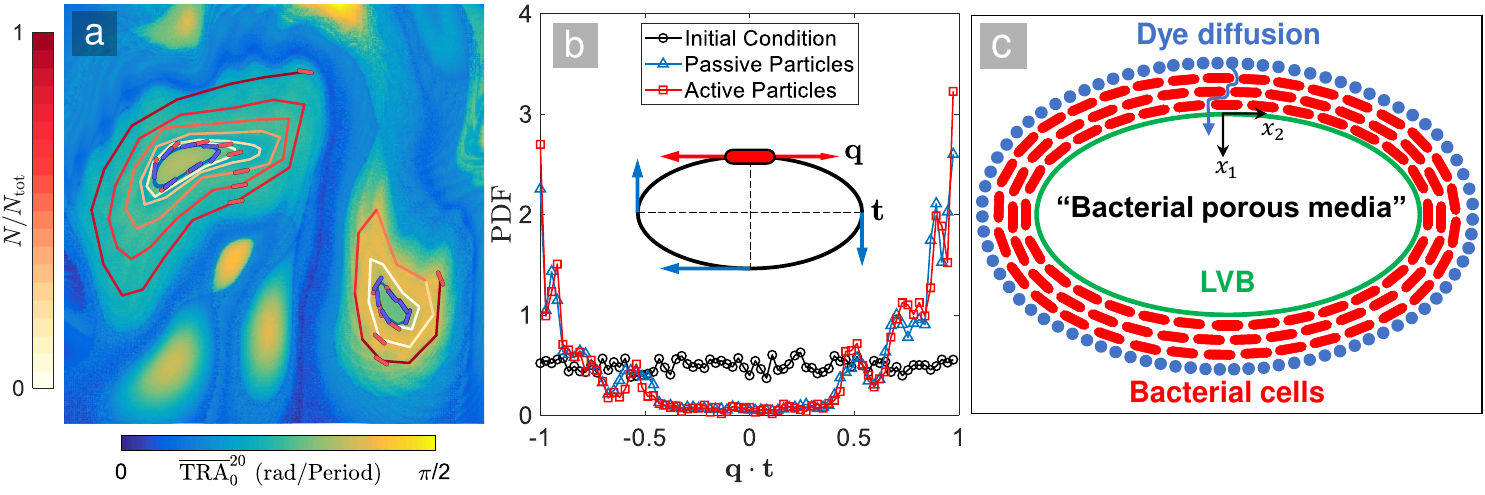}}
    \caption{(\textit{a}) Stroboscopic trajectories of passive (blue) and active (red) particles that are initially inside the Lagrangian vortices illustrated by the TRA field. While sharing the same initial condition, passive particles remain trapped in the vortices and active particles spiral outward and escape. The trajectories of active particles are colored by their normalized time (or numbers of periods), $N/N_{\mathrm{tot}}$, where $N_{\mathrm{tot}}$ is the total time duration of the trajectories. (\textit{b}) The PDFs of the inner product of the particle orientation vector, $\mathbf{q}$, and the tangent vector of the elliptic LCSs in the direction of the vortex circulation, $\mathbf{t}$, as defined in the inset. The initial condition ($N=0$) is approximately a uniform distribution for particles with random initial orientations. The PDFs at a later time ($N=50$) show that both passive and active particles preferentially align with members of elliptic LCSs at $\mathbf{q}\cdot\mathbf{t}=\pm 1$. The PDF of active particles is slightly biased towards the positive peak of $\mathbf{q}\cdot\mathbf{t}=+1$. (\textit{c}) Schematic of a ``bacterial porous medium'' formed by cells aligning and accumulating outside a Lagrangian vortex boundary (LVB). Dye transport is hindered as it diffuses through the porous media.}
\label{fig4}
\end{figure}

To gain further insights into how swimming particles are repelled by Lagrangian vortices, we plot the stroboscopic trajectories and orientations of passive and active particles as a function of time, as shown in Fig. \hyperref[fig4]{4(a)}. The Lagrangian vortices are illustrated by the TRA field (color map); passive and active particles are shown as blue and red bars, respectively. The trajectories of the passive particles are blue, while the trajectories of the active particles are colored by their normalized time or number of periods, $N/N_\mathrm{tot}$, where $N_\mathrm{tot}$ is the total time duration of the trajectories. Despite sharing the same initial conditions for position and orientation, passive particles remain trapped in the Lagrangian vortices while active particles spiral outward with time and escape the vortices. We note that the swimming number of the active particles, $\Phi=v_s/U$, is of the order of $10^{-2}$ in the simulations, suggesting that the swimming speed is fairly negligible compared to the flow speed. However, even such a (relatively) low swimming speed can drive active particles out of equilibrium and escape the Lagrangian vortices. That is, even small levels of swimming activity is enough to initiate the expulsion process. In addition, we find that the orientations of both passive and active particles tend to align with the level curves of the TRA field. Because the convex TRA contours locate the nested families of elliptic LCSs, this result indicates elongated particles---whether self-propelled or not---preferentially align with the members of elliptic LCSs. The difference is that the self-propulsion of active particles causes them to move outward and leave the vortices in a spiral manner.

Next, we test the alignment of the elongated particles by calculating the inner product of the particle orientation vector, $\mathbf{q}$, and the tangent vector of the elliptic LCSs in the direction of vortex circulation, $\mathbf{t}$. The inset of Fig. \hyperref[fig4]{4(b)} shows a schematic of the two vectors. Since elliptic LCSs are nested families, we partition the elliptic LCSs into 10 members using the convex TRA contours. The inner product is calculated between the orientation vector of an elongated particle and the tangent of the nearest member of elliptic LCSs. Figure \hyperref[fig4]{4(b)} shows probability density functions (PDFs) of $\mathbf{q}\cdot\mathbf{t}$ for both passive and active particles at two different times. At $N=0$, the initial condition of the PDFs shared by both passive and active particles is approximately a uniform distribution due to the random initial orientations. At a later time of $N=50$, the PDFs of both passive and active particles become bimodal at $\mathbf{q}\cdot\mathbf{t}=\pm 1$, suggesting a parallel or anti-parallel alignment between $\mathbf{q}$ and $\mathbf{t}$. We find that the PDF of active particles shows a slight bias towards the positive peak at $\mathbf{q}\cdot\mathbf{t}=+1$, while the bias is not present for the PDF of passive particles. This suggests that active particles prefer to swim parallel to the elliptic LCSs in the direction of vortex circulation rather than swimming against the circulation. The bias in parallel and anti-parallel alignments is also found for active particles of different swimming speeds (see Supplementary Materials). The self-propulsion of active particles can cause them to travel toward the outer members of the elliptic LCSs in both parallel and anti-parallel alignment configurations (see Movie S4), resulting in the spiral escaping trajectories of the active particles shown in Fig. \hyperref[fig4]{4(a)}. Consequently, the interplay between self-propulsion and the alignment of elongated particles with elliptic LCSs leads to the depletion of active particles inside the elliptic LCSs, that is, the accumulation of active particles outside the Lagrangian vortex boundaries (LVBs). This accumulation of microswimmers outside the LVBs can further obstruct dye transport into the Lagrangian vortices, which is responsible for the observed transport barriers.

\subsection{Physical Mechanism}
We now propose a potential mechanism to explain the observed reduction in dye transport into Lagrangian vortices. We posit that the alignment and accumulation of bacteria outside the LVBs form an effective ``bacterial porous medium'' that can impede the diffusion of dye molecules, as schematically shown in Fig. \hyperref[fig4]{4(c)}. In the absence of bacteria, dye should enter LVBs primarily by diffusion and marginally by inertial effect due to (low but) nonzero Reynolds number ($Re\sim10^1$). Here, we will focus on diffusive transport around the outside boundary of the LVBs where bacteria accumulate [Fig. \hyperref[fig4]{4(c)}]. The effective diffusivity of a molecular dye through a porous medium, $D_{\mathrm{eff}}$, can be estimated as follows \citep{Grathwohl1998}:
\begin{equation}\label{eqn. porous diffusivity}
    D_{\mathrm{eff}} = \frac{D_0\epsilon\delta}{\tau},
\end{equation}
where $D_0$ is the intrinsic diffusivity of the dye molecules, $\epsilon$, $\delta$, and $\tau$ are the porosity, constrictivity, and tortuosity of the porous media, respectively. The porosity, $\epsilon$, is the volume fraction of the void spaces (that is, pores) in the medium relative to its total volume; $\epsilon$ ranges from 0 to 1. For our bacterial medium, the quantity $\epsilon$ sums up to unity with the bacterial volume fraction $\phi_b$ such that $\epsilon=1-\phi_b$. Bacterial accumulation around LVB leads to a local increase in $\phi_b$ of approximately tenfold (as shown by our simulations) such that $\phi_{\mathrm{LVB}}\approx 10\phi_b=0.05$. Thus, the porosity of the bacterial medium is $\epsilon = 1-\phi_{\mathrm{LVB}}\approx0.95$. This leads to only a 5\% decrease in $D_{\mathrm{eff}}$, which cannot account for the observed reduction in dye transport. Similarly, the constrictivity $\delta$ is a dimensionless parameter ranging from 0 to 1, which captures the hindrance to which a diffusing substance is subjected to when travelling through narrow pores. It becomes important only if the size of the diffusing molecules is comparable to that of the pores \citep{Stenzel_2016,Bini2019}. Here, the pore size, $\mathcal{O}(1~\umu$m), is much greater than the molecular size, $\mathcal{O}$(1 nm), and therefore $\delta\approx1$. Since the values of $\delta$ and $\epsilon$ are close to unity, we do not expect them to play a significant role in the reduction of dye transport.

Next, we examine the role of tortuosity ($\tau$), which compares the (tortuous) pathway of molecular diffusion in a porous medium to its pathway in an unrestricted medium \citep{Bini2019,daSilva_2022}. The quantity $\tau$ can be defined as \citep{Grathwohl1998,Holzer2013,daSilva_2022}:
\begin{equation}
     \tau = \frac{\langle L_s\rangle}{L_0},
\end{equation}
where $L_0$ is the length of the straightest path and $\langle L_s\rangle$ is the ensemble average of all possible tortuous path of diffusion. Note that $\tau\ge1$ since $\langle L_s\rangle \ge L_0$. We consider the transverse diffusion of dye molecules across a bacterial porous medium, as sketched in Fig. \hyperref[fig4]{4(c)}, where $x_1$ denotes the direction of the transverse diffusion and $x_2$ denotes the direction of cell body alignment with the tangent of the elliptic LCSs. During the diffusion process, each time the (dye) molecule encounters an obstacle (i.e., a bacterium), its path will be deflected by on average half the body length of a bacterium, $l_E/2$. Also, dye diffusing a distance $L_0$ (in the bacterial porous medium) will encounter bacterial cells on average $L_0\sigma_N$ times, where $\sigma_N$ is the cell number density. We can then estimate the average transverse and longitudinal path-lengths, $\langle L_s\rangle_{11}$ and $\langle L_s\rangle_{22}$, for the dye to diffuse across a distance $L_0$ in the $x_1$ and $x_2$ directions as:
\begin{equation}
    \langle L_s\rangle_{11} = L_0+L_0\sigma_Nl_E/2,\ \langle L_s\rangle_{22} = L_0+L_0\sigma_N \alpha l_E/2,
\end{equation}
where $\alpha = d_E/l_E$ is the aspect ratio between the average cell diameter $d_E$ and cell length $l_E$. For rod-shaped bacteria such as \textit{E. coli}, we notice that $\alpha<1$ and thus $\langle L_s\rangle_{22}<\langle L_s\rangle_{11}$. This suggests that diffusion is anisotropic in the bacterial porous medium. We can now express the tortuosity for the transverse and longitudinal diffusion as:
\begin{equation}\label{eqn. tortuosity}
    \tau_{11} =  \frac{\langle L_s\rangle_{11}}{L_0}= 1+\sigma_N l_E/2,\ \tau_{22} =  \frac{\langle L_s\rangle_{22}}{L_0}= 1+\alpha\sigma_N l_E/2.
\end{equation}

In our experiments, the local bacterial volume fraction (around LVBs) is $\phi_{\mathrm{LVB}}\approx 0.05$, which corresponds to a local cell number density of $\rho_N \approx 1.25\times10^{11}$ cells/mL, and a local cell number density of $\sigma_N \approx 5,000$ cells/cm. If we consider $l_E$ to be the bacterium body length ($\approx2~\umu$m), then $\tau_{11} = 1+\sigma_N l_E/2=1.5$. If instead we consider the "total" bacterium length (cell body plus flagella), $L_{\mathrm{total}}\approx7~\umu$m \citep{Patteson2015},  then $\tau_{11} = 1+\sigma_N L_{\mathrm{total}}/2=2.75$. Using the above estimates, we expect $1.5\le\tau_{11}\le2.75$. The transverse effective diffusivity can be estimated using Eq. \ref{eqn. porous diffusivity}, and this yields $0.346\le D_{11}/D_0\le0.633$. Therefore, a dye diffusing through this bacterial porous media would experience a 33\%--66\% decrease in the transverse effective diffusivity, which qualitatively corresponds to the decrease in dye transport into the LVBs. Note that Eq. \ref{eqn. tortuosity} estimates an anisotropic diffusivity, $D_{11}<D_{22}$, for rod-shaped bacteria ($\alpha<1$). Bacterial flagellar motion, in addition, can displace dye molecules further in the $x_2$ direction, which would lead to an increase in $D_{22}$. This would lead to a more complex behavior than the one described here. Nevertheless, the concept of anisotropic diffusion in a bacterial porous media seems to capture, at least qualitatively, the observed decrease in scalar transport into the Lagrangian vortices.

\section{Conclusion}
In this manuscript, we investigate the interaction between swimming microorganisms and Lagrangian coherent vortices known as the elliptic LCSs in time-periodic flows in experiments and in simulations. Our results show that even small amounts of swimming activity can affect (i) the dynamics of active particles in the flow (Fig.~\ref{fig4}) and consequently (ii) the mixing and transport of passive scalars (Fig.~\ref{fig1}) in chaotic flows. Experiments show that the interaction between organisms and elliptic LCSs leads to transport barriers through which the tracer flux is significantly reduced. Using the Poincar\'e map and the TRA field, we show that these transport barriers coincide with outermost member of elliptic LCSs, or Lagrangian vortex boundaries (LVBs). To further understand the formation of the transport barriers, we perform numerical simulations of elongated microswimmers in experimentally measured velocity fields. Results show that elliptic LCSs can repel elongated swimmer and lead to swimmer accumulation outside LVBs. This accumulation of microswimmers effectively reduces the transport into elliptic LCSs. We further show that the interplay between self-propulsion and the preferential alignment of elongated particle with the tangents of elliptic LCSs leads swimmers to escape the Lagrangian vortices. Overall, our results allow quantitative prediction of the Lagrangian transport of microorganisms and passive tracer quantities (e.g., temperature, oxygen, and nutrients) in chaotic flows with nontrivial vortex structures. Although there have been previous studies on the interaction between microorganisms and LCSs \citep{Khurana_PoF_2012,Dehkharghani_PNAS_2019,Ran_PNAS_2021,Qin_Sci_Adv_2022,Si_Fang_PoF_2021,Si_Fang_PRF_2022,Solomon_SwIMs_2022},  our work extend study of microorganism-LCSs interaction to elliptic LCSs (i.e., vortex-like dynamical structures). From a practical perspective, our results may be useful in understanding the organic matter flux in the oceans, algal blooms in lakes, and harmful bacterial infections. 

\section*{Acknowledgement}
We thank Justin C. Burton, Tom Solomon, Kevin Mitchell, Simon Berman, Douglas Jerolmack, and Albane Th\'ery for insightful discussions, and Brendan Blackwell for help with early work. This work was supported by NSF-DMR-1709763.

\bibliographystyle{jfm}

\bibliography{reference}

\begin{thebibliography}{58}
\expandafter\ifx\csname natexlab\endcsname\relax\def\natexlab#1{#1}\fi
\def\au#1{#1} \def\ed#1{#1} \def\yr#1{#1}\def\at#1{#1}\def\jt#1{\textit{#1}} \def\bt#1{#1}\def\bvol#1{\textbf{#1}} \def\vol#1{#1} \def\pg#1{#1} \def\publ#1{#1}\def\arxiv#1{#1}\def\org#1{#1}\def\st#1{\textit{#1}}

\bibitem[Aksamit \& Haller(2022)]{Aksamit_Haller_JFM_2022}
{\sc \au{Aksamit, Nikolas~O.} \& \au{Haller, George}} \yr{2022}  \at{Objective momentum barriers in wall turbulence}.  \jt{J. Fluid Mech.}  \bvol{941},  \pg{A3}.

\bibitem[Barry {\em et~al.\/}(2015)Barry, Rusconi, Guasto \& Stocker]{Stocker_JRSI_2015}
{\sc \au{Barry, Michael~T.}, \au{Rusconi, Roberto}, \au{Guasto, Jeffrey~S.} \& \au{Stocker, Roman}} \yr{2015}  \at{Shear-induced orientational dynamics and spatial heterogeneity in suspensions of motile phytoplankton}.  \jt{J. R. Soc. Interface}  \bvol{12}~(112),  \pg{20150791}.

\bibitem[Bini {\em et~al.\/}(2019)Bini, Pica, Marinozzi \& Marinozzi]{Bini2019}
{\sc \au{Bini, Fabiano}, \au{Pica, Andrada}, \au{Marinozzi, Andrea} \& \au{Marinozzi, Franco}} \yr{2019}  \at{A {3D} model of the effect of tortuosity and constrictivity on the diffusion in mineralized collagen fibril}.  \jt{Sci. Rep.}  \bvol{9}~(1),  \pg{2658}.

\bibitem[Borgnino {\em et~al.\/}(2018)Borgnino, Boffetta, De~Lillo \& Cencini]{Borgnino_JFM_2018}
{\sc \au{Borgnino, M.}, \au{Boffetta, G.}, \au{De~Lillo, F.} \& \au{Cencini, M.}} \yr{2018}  \at{Gyrotactic swimmers in turbulence: shape effects and role of the large-scale flow}.  \jt{J. Fluid Mech.}  \bvol{856},  \pg{R1}.

\bibitem[Borgnino {\em et~al.\/}(2019)Borgnino, Gustavsson, De~Lillo, Boffetta, Cencini \& Mehlig]{Borgnino_PRL_2019}
{\sc \au{Borgnino, M.}, \au{Gustavsson, K.}, \au{De~Lillo, F.}, \au{Boffetta, G.}, \au{Cencini, M.} \& \au{Mehlig, B.}} \yr{2019}  \at{Alignment of nonspherical active particles in chaotic flows}.  \jt{Phys. Rev. Lett.}  \bvol{123},  \pg{138003}.

\bibitem[Costerton {\em et~al.\/}(1999)Costerton, Stewart \& Greenberg]{Costerton_Science_1999}
{\sc \au{Costerton, J.~W.}, \au{Stewart, Philip~S.} \& \au{Greenberg, E.~P.}} \yr{1999}  \at{Bacterial biofilms: A common cause of persistent infections}.  \jt{Science}  \bvol{284}~(5418),  \pg{1318--1322}.

\bibitem[Crocker \& Grier(1996)]{Crocker_1996}
{\sc \au{Crocker, John~C.} \& \au{Grier, David~G.}} \yr{1996}  \at{Methods of digital video microscopy for colloidal studies}.  \jt{J. Colloid Interface Sci.}  \bvol{179}~(1),  \pg{298--310}.

\bibitem[De~Lillo {\em et~al.\/}(2014)De~Lillo, Cencini, Durham, Barry, Stocker, Climent \& Boffetta]{DeLillo_PRL_2014}
{\sc \au{De~Lillo, Filippo}, \au{Cencini, Massimo}, \au{Durham, William~M.}, \au{Barry, Michael}, \au{Stocker, Roman}, \au{Climent, Eric} \& \au{Boffetta, Guido}} \yr{2014}  \at{Turbulent fluid acceleration generates clusters of gyrotactic microorganisms}.  \jt{Phys. Rev. Lett.}  \bvol{112},  \pg{044502}.

\bibitem[Dehkharghani {\em et~al.\/}(2019)Dehkharghani, Waisbord, Dunkel \& Guasto]{Dehkharghani_PNAS_2019}
{\sc \au{Dehkharghani, A.}, \au{Waisbord, N.}, \au{Dunkel, J.} \& \au{Guasto, J.~S.}} \yr{2019}  \at{Bacterial scattering in microfluidic crystal flows reveals giant active taylor{\textendash}aris dispersion}.  \jt{Proc. Natl. Acad. Sci. U.S.A.}  \bvol{116}~(23),  \pg{11119--11124}.

\bibitem[Durham {\em et~al.\/}(2013)Durham, Climent, Barry, De~Lillo, Boffetta, Cencini \& Stocker]{Durham_NC_2013}
{\sc \au{Durham, William~M.}, \au{Climent, Eric}, \au{Barry, Michael}, \au{De~Lillo, Filippo}, \au{Boffetta, Guido}, \au{Cencini, Massimo} \& \au{Stocker, Roman}} \yr{2013}  \at{Turbulence drives microscale patches of motile phytoplankton}.  \jt{Nat. Commun.}  \bvol{4}~(1),  \pg{2148}.

\bibitem[Durham {\em et~al.\/}(2009)Durham, Kessler \& Stocker]{Durham_Science_2009}
{\sc \au{Durham, William~M.}, \au{Kessler, John~O.} \& \au{Stocker, Roman}} \yr{2009}  \at{Disruption of vertical motility by shear triggers formation of thin phytoplankton layers}.  \jt{Science}  \bvol{323}~(5917),  \pg{1067--1070}.

\bibitem[Ezhilan \& Saintillan(2015)]{Saintillan_JFM_2015}
{\sc \au{Ezhilan, B.} \& \au{Saintillan, D.}} \yr{2015}  \at{Transport of a dilute active suspension in pressure-driven channel flow}.  \jt{J. Fluid Mech.}  \bvol{777},  \pg{482–522}.

\bibitem[Farazmand \& Haller(2016)]{Farazmand_PhysD_2016}
{\sc \au{Farazmand, Mohammad} \& \au{Haller, George}} \yr{2016}  \at{Polar rotation angle identifies elliptic islands in unsteady dynamical systems}.  \jt{Physica D}  \bvol{315},  \pg{1--12}.

\bibitem[Grathwohl(1998)]{Grathwohl1998}
{\sc \au{Grathwohl, Peter}} \yr{1998} {\em Diffusion in Natural Porous Media: Contaminant Transport, Sorption/Desorption and Dissolution Kinetics\/},  \pg{pp. 43--81}.  \publ{Boston, MA: Springer US}.

\bibitem[Guasto {\em et~al.\/}(2012)Guasto, Rusconi \& Stocker]{Guasto_Annu_Rev_2012}
{\sc \au{Guasto, Jeffrey~S.}, \au{Rusconi, Roberto} \& \au{Stocker, Roman}} \yr{2012}  \at{Fluid mechanics of planktonic microorganisms}.  \jt{Annu. Rev. Fluid Mech.}  \bvol{44}~(1),  \pg{373--400}.

\bibitem[Gustavsson {\em et~al.\/}(2016)Gustavsson, Berglund, Jonsson \& Mehlig]{Gustavsson_PRL_2016}
{\sc \au{Gustavsson, K.}, \au{Berglund, F.}, \au{Jonsson, P.~R.} \& \au{Mehlig, B.}} \yr{2016}  \at{Preferential sampling and small-scale clustering of gyrotactic microswimmers in turbulence}.  \jt{Phys. Rev. Lett.}  \bvol{116},  \pg{108104}.

\bibitem[Haller(2015)]{Haller_ARFM_2015}
{\sc \au{Haller, G.}} \yr{2015}  \at{Lagrangian coherent structures}.  \jt{Annu. Rev. Fluid Mech.}  \bvol{47}~(1),  \pg{137--162}.

\bibitem[Haller {\em et~al.\/}(2021)Haller, Aksamit \& Encinas-Bartos]{Haller_Chaos_TRA_2021}
{\sc \au{Haller, George}, \au{Aksamit, Nikolas} \& \au{Encinas-Bartos, Alex~P.}} \yr{2021}  \at{Quasi-objective coherent structure diagnostics from single trajectories}.  \jt{Chaos}  \bvol{31}~(4),  \pg{043131}.

\bibitem[Haller {\em et~al.\/}(2016)Haller, Hadjighasem, Farazmand \& Huhn]{Haller_JFM_2016}
{\sc \au{Haller, G.}, \au{Hadjighasem, A.}, \au{Farazmand, M.} \& \au{Huhn, F.}} \yr{2016}  \at{Defining coherent vortices objectively from the vorticity}.  \jt{J. Fluid Mech.}  \bvol{795},  \pg{136–173}.

\bibitem[Haller {\em et~al.\/}(2020)Haller, Katsanoulis, Holzner, Frohnapfel \& Gatti]{Haller_JFM_2020}
{\sc \au{Haller, George}, \au{Katsanoulis, Stergios}, \au{Holzner, Markus}, \au{Frohnapfel, Bettina} \& \au{Gatti, Davide}} \yr{2020}  \at{Objective barriers to the transport of dynamically active vector fields}.  \jt{J. Fluid Mech.}  \bvol{905},  \pg{A17}.

\bibitem[Hill {\em et~al.\/}(2007)Hill, Kalkanci, McMurry \& Koser]{Hill_PRL_2007}
{\sc \au{Hill, Jane}, \au{Kalkanci, Ozge}, \au{McMurry, Jonathan~L.} \& \au{Koser, Hur}} \yr{2007}  \at{Hydrodynamic surface interactions enable escherichia coli to seek efficient routes to swim upstream}.  \jt{Phys. Rev. Lett.}  \bvol{98},  \pg{068101}.

\bibitem[Holzer {\em et~al.\/}(2013)Holzer, Wiedenmann, M{\"u}nch, Keller, Prestat, Gasser, Robertson \& Grob{\'e}ty]{Holzer2013}
{\sc \au{Holzer, L.}, \au{Wiedenmann, D.}, \au{M{\"u}nch, B.}, \au{Keller, L.}, \au{Prestat, M.}, \au{Gasser, Ph.}, \au{Robertson, I.} \& \au{Grob{\'e}ty, B.}} \yr{2013}  \at{The influence of constrictivity on the effective transport properties of porous layers in electrolysis and fuel cells}.  \jt{J. Mater. Sci.}  \bvol{48}~(7),  \pg{2934--2952}.

\bibitem[Jeffery \& Filon(1922)]{Jeffery1922}
{\sc \au{Jeffery, George~Barker} \& \au{Filon, Louis Napoleon~George}} \yr{1922}  \at{The motion of ellipsoidal particles immersed in a viscous fluid}.  \jt{Proc. R. Soc. Lond. A}  \bvol{102}~(715),  \pg{161--179}.

\bibitem[Kasyap {\em et~al.\/}(2014)Kasyap, Koch \& Wu]{Kasyap_PoF_2014}
{\sc \au{Kasyap, T.~V.}, \au{Koch, Donald~L.} \& \au{Wu, Mingming}} \yr{2014}  \at{Hydrodynamic tracer diffusion in suspensions of swimming bacteria}.  \jt{Phys. Fluids}  \bvol{26}~(8),  \pg{081901}.

\bibitem[Katsanoulis {\em et~al.\/}(2020)Katsanoulis, Farazmand, Serra \& Haller]{Katsanoulis_PRF_2020}
{\sc \au{Katsanoulis, Stergios}, \au{Farazmand, Mohammad}, \au{Serra, Mattia} \& \au{Haller, George}} \yr{2020}  \at{Vortex boundaries as barriers to diffusive vorticity transport in two-dimensional flows}.  \jt{Phys. Rev. Fluids}  \bvol{5},  \pg{024701}.

\bibitem[Khurana {\em et~al.\/}(2011)Khurana, Blawzdziewicz \& Ouellette]{Khurana_PRL_2011}
{\sc \au{Khurana, Nidhi}, \au{Blawzdziewicz, Jerzy} \& \au{Ouellette, Nicholas~T.}} \yr{2011}  \at{Reduced transport of swimming particles in chaotic flow due to hydrodynamic trapping}.  \jt{Phys. Rev. Lett.}  \bvol{106},  \pg{198104}.

\bibitem[Khurana \& Ouellette(2012)]{Khurana_PoF_2012}
{\sc \au{Khurana, Nidhi} \& \au{Ouellette, Nicholas~T.}} \yr{2012}  \at{Interactions between active particles and dynamical structures in chaotic flow}.  \jt{Phys. Fluids}  \bvol{24}~(9),  \pg{091902}.

\bibitem[Kim {\em et~al.\/}(2016)Kim, Ingremeau, Zhao, Bassler \& Stone]{Kim_Nat_MicroBio_2016}
{\sc \au{Kim, Minyoung~Kevin}, \au{Ingremeau, Fran{\c{c}}ois}, \au{Zhao, Aishan}, \au{Bassler, Bonnie~L.} \& \au{Stone, Howard~A.}} \yr{2016}  \at{Local and global consequences of flow on bacterial quorum sensing}.  \jt{Nat. Microbiol.}  \bvol{1}~(1),  \pg{15005}.

\bibitem[Locsei \& Pedley(2009)]{Locsei_BMB_2009}
{\sc \au{Locsei, J.~T.} \& \au{Pedley, T.~J.}} \yr{2009}  \at{Run and tumble chemotaxis in a shear flow: The effect of temporal comparisons, persistence, rotational diffusion, and cell shape}.  \jt{Bull. Math. Biol.}  \bvol{71}~(5),  \pg{1089--1116}.

\bibitem[Marcos {\em et~al.\/}(2012)Marcos, Fu, Powers \& Stocker]{Marcos_PNAS_2012}
{\sc \au{Marcos}, \au{Fu, Henry~C.}, \au{Powers, Thomas~R.} \& \au{Stocker, Roman}} \yr{2012}  \at{Bacterial rheotaxis}.  \jt{Proc. Natl. Acad. Sci. U.S.A.}  \bvol{109}~(13),  \pg{4780--4785}.

\bibitem[Mathijssen {\em et~al.\/}(2019)Mathijssen, Figueroa-Morales, Junot, Cl{\'e}ment, Lindner \& Z{\"o}ttl]{Mathijssen_NC_2019}
{\sc \au{Mathijssen, A. J. T.~M.}, \au{Figueroa-Morales, N.}, \au{Junot, G.}, \au{Cl{\'e}ment, E.}, \au{Lindner, A.} \& \au{Z{\"o}ttl, A.}} \yr{2019}  \at{Oscillatory surface rheotaxis of swimming \textit{E. coli} bacteria}.  \jt{Nat. Commun.}  \bvol{10}~(1),  \pg{3434}.

\bibitem[Musielak {\em et~al.\/}(2009)Musielak, Karp-Boss, Jumars \& Fauci]{Musielak_JFM_2009}
{\sc \au{Musielak, M.~M.}, \au{Karp-Boss, Lee}, \au{Jumars, P.~A.} \& \au{Fauci, L.~J.}} \yr{2009}  \at{Nutrient transport and acquisition by diatom chains in a moving fluid}.  \jt{J. Fluid Mech.}  \bvol{638},  \pg{401–421}.

\bibitem[Ottino(1989)]{Ottino1989}
{\sc \au{Ottino, J.M.}} \yr{1989} {\em The Kinematics of Mixing: Stretching, Chaos, and Transport\/}.  \publ{New York: Cambridge University Press}.

\bibitem[Parsa {\em et~al.\/}(2011)Parsa, Guasto, Kishore, Ouellette, Gollub \& Voth]{Parsa_PoF_2011}
{\sc \au{Parsa, Shima}, \au{Guasto, Jeffrey~S.}, \au{Kishore, Monica}, \au{Ouellette, Nicholas~T.}, \au{Gollub, J.~P.} \& \au{Voth, Greg~A.}} \yr{2011}  \at{{Rotation and alignment of rods in two-dimensional chaotic flow}}.  \jt{Phys. Fluids}  \bvol{23}~(4),  \pg{043302}.

\bibitem[Patteson {\em et~al.\/}(2015)Patteson, Gopinath, Goulian \& Arratia]{Patteson2015}
{\sc \au{Patteson, A.~E.}, \au{Gopinath, A.}, \au{Goulian, M.} \& \au{Arratia, P.~E.}} \yr{2015}  \at{Running and tumbling with {E. coli} in polymeric solutions}.  \jt{Sci. Rep.}  \bvol{5}~(1),  \pg{15761}.

\bibitem[Pujara {\em et~al.\/}(2018)Pujara, Koehl \& Variano]{Pujara_JFM_2018}
{\sc \au{Pujara, N.}, \au{Koehl, M. A.~R.} \& \au{Variano, E.~A.}} \yr{2018}  \at{Rotations and accumulation of ellipsoidal microswimmers in isotropic turbulence}.  \jt{J. Fluid Mech.}  \bvol{838},  \pg{356–368}.

\bibitem[Qin \& Arratia(2022)]{Qin_Sci_Adv_2022}
{\sc \au{Qin, Boyang} \& \au{Arratia, Paulo~E.}} \yr{2022}  \at{Confinement, chaotic transport, and trapping of active swimmers in time-periodic flows}.  \jt{Sci. Adv.}  \bvol{8}~(49),  \pg{eadd6196}.

\bibitem[Ran {\em et~al.\/}(2021)Ran, Brosseau, Blackwell, Qin, Winter \& Arratia]{Ran_PNAS_2021}
{\sc \au{Ran, Ranjiangshang}, \au{Brosseau, Quentin}, \au{Blackwell, Brendan~C.}, \au{Qin, Boyang}, \au{Winter, Rebecca~L.} \& \au{Arratia, Paulo~E.}} \yr{2021}  \at{Bacteria hinder large-scale transport and enhance small-scale mixing in time-periodic flows}.  \jt{Proc. Natl. Acad. Sci. U.S.A.}  \bvol{118},  \pg{e2108548118}.

\bibitem[Ran {\em et~al.\/}(2022)Ran, Brosseau, Blackwell, Qin, Winter \& Arratia]{Ran_PRF_2022}
{\sc \au{Ran, Ranjiangshang}, \au{Brosseau, Quentin}, \au{Blackwell, Brendan~C.}, \au{Qin, Boyang}, \au{Winter, Rebecca~L.} \& \au{Arratia, Paulo~E.}} \yr{2022}  \at{Mixing in chaotic flows with swimming bacteria}.  \jt{Phys. Rev. Fluids}  \bvol{7},  \pg{110511}.

\bibitem[Riffell \& Zimmer(2007)]{Zimmer_J_Exp_Biol_2007}
{\sc \au{Riffell, Jeffrey~A.} \& \au{Zimmer, Richard~K.}} \yr{2007}  \at{Sex and flow: the consequences of fluid shear for sperm–egg interactions}.  \jt{J. Exp. Biol.}  \bvol{210}~(20),  \pg{3644--3660}.

\bibitem[Rusconi {\em et~al.\/}(2014)Rusconi, Guasto \& Stocker]{Rusconi_NatPhys_2014}
{\sc \au{Rusconi, Roberto}, \au{Guasto, Jeffrey~S.} \& \au{Stocker, Roman}} \yr{2014}  \at{Bacterial transport suppressed by fluid shear}.  \jt{Nat. Phys.}  \bvol{10}~(3),  \pg{212--217}.

\bibitem[Si \& Fang(2021)]{Si_Fang_PoF_2021}
{\sc \au{Si, Xinyu} \& \au{Fang, Lei}} \yr{2021}  \at{{Preferential alignment and heterogeneous distribution of active non-spherical swimmers near Lagrangian coherent structures}}.  \jt{Phys. Fluids}  \bvol{33}~(7),  \pg{073303}.

\bibitem[Si \& Fang(2022)]{Si_Fang_PRF_2022}
{\sc \au{Si, Xinyu} \& \au{Fang, Lei}} \yr{2022}  \at{Preferential transport of swimmers in heterogeneous two-dimensional turbulent flow}.  \jt{Phys. Rev. Fluids}  \bvol{7},  \pg{094501}.

\bibitem[da~Silva {\em et~al.\/}(2022)da~Silva, do~Rocio~Cardoso, Pabst~Veronese \& Mazer]{daSilva_2022}
{\sc \au{da~Silva, Marly Terezinha Quadri Sim\~oes}, \au{do~Rocio~Cardoso, Marianna}, \au{Pabst~Veronese, Caterina~Maria} \& \au{Mazer, Wellington}} \yr{2022}  \at{Tortuosity: A brief review}.  \jt{Mater. Today Proc.}  \bvol{58},  \pg{1344--1349}.

\bibitem[Stenzel {\em et~al.\/}(2016)Stenzel, Pecho, Holzer, Neumann \& Schmidt]{Stenzel_2016}
{\sc \au{Stenzel, Ole}, \au{Pecho, Omar}, \au{Holzer, Lorenz}, \au{Neumann, Matthias} \& \au{Schmidt, Volker}} \yr{2016}  \at{Predicting effective conductivities based on geometric microstructure characteristics}.  \jt{AIChE J.}  \bvol{62}~(5),  \pg{1834--1843}.

\bibitem[Stocker \& Seymour(2012)]{Stocker_MMBR_2012}
{\sc \au{Stocker, Roman} \& \au{Seymour, Justin~R.}} \yr{2012}  \at{Ecology and physics of bacterial chemotaxis in the ocean}.  \jt{Microbiol. Mol. Biol. Rev.}  \bvol{76}~(4),  \pg{792--812}.

\bibitem[Stocker {\em et~al.\/}(2008)Stocker, Seymour, Samadani, Hunt \& Polz]{Stocker_PNAS_2008}
{\sc \au{Stocker, Roman}, \au{Seymour, Justin~R.}, \au{Samadani, Azadeh}, \au{Hunt, Dana~E.} \& \au{Polz, Martin~F.}} \yr{2008}  \at{Rapid chemotactic response enables marine bacteria to exploit ephemeral microscale nutrient patches}.  \jt{Proc. Natl. Acad. Sci. U.S.A.}  \bvol{105}~(11),  \pg{4209--4214}.

\bibitem[Taylor \& Stocker(2012)]{Taylor_Science_2012}
{\sc \au{Taylor, John~R.} \& \au{Stocker, Roman}} \yr{2012}  \at{Trade-offs of chemotactic foraging in turbulent water}.  \jt{Science}  \bvol{338}~(6107),  \pg{675--679}.

\bibitem[Torney \& Neufeld(2007)]{Torney_PRL_2007}
{\sc \au{Torney, Colin} \& \au{Neufeld, Zolt\'an}} \yr{2007}  \at{Transport and aggregation of self-propelled particles in fluid flows}.  \jt{Phys. Rev. Lett.}  \bvol{99},  \pg{078101}.

\bibitem[Tung {\em et~al.\/}(2015)Tung, Ardon, Roy, Koch, Suarez \& Wu]{Tung_PRL_2015}
{\sc \au{Tung, Chih-Kuan}, \au{Ardon, Florencia}, \au{Roy, Anubhab}, \au{Koch, Donald~L.}, \au{Suarez, Susan~S.} \& \au{Wu, Mingming}} \yr{2015}  \at{Emergence of upstream swimming via a hydrodynamic transition}.  \jt{Phys. Rev. Lett.}  \bvol{114},  \pg{108102}.

\bibitem[Voth {\em et~al.\/}(2002)Voth, Haller \& Gollub]{Voth_PRL_2002}
{\sc \au{Voth, Greg~A.}, \au{Haller, G.} \& \au{Gollub, J.~P.}} \yr{2002}  \at{Experimental measurements of stretching fields in fluid mixing}.  \jt{Phys. Rev. Lett.}  \bvol{88},  \pg{254501}.

\bibitem[Voth {\em et~al.\/}(2003)Voth, Saint, Dobler \& Gollub]{Voth_PoF_2003}
{\sc \au{Voth, Greg~A.}, \au{Saint, T.~C.}, \au{Dobler, Greg} \& \au{Gollub, J.~P.}} \yr{2003}  \at{Mixing rates and symmetry breaking in two-dimensional chaotic flow}.  \jt{Phys. Fluids}  \bvol{15}~(9),  \pg{2560--2566}.

\bibitem[Wheeler {\em et~al.\/}(2019)Wheeler, Secchi, Rusconi \& Stocker]{Stocker_Annu_Rev_2019}
{\sc \au{Wheeler, Jeanette~D.}, \au{Secchi, Eleonora}, \au{Rusconi, Roberto} \& \au{Stocker, Roman}} \yr{2019}  \at{Not just going with the flow: The effects of fluid flow on bacteria and plankton}.  \jt{Annu. Rev. Cell Dev. Biol.}  \bvol{35}~(1),  \pg{213--237}.

\bibitem[Yawata {\em et~al.\/}(2016)Yawata, Nguyen, Stocker \& Rusconi]{Stocker_JBac_2016}
{\sc \au{Yawata, Yutaka}, \au{Nguyen, Jen}, \au{Stocker, Roman} \& \au{Rusconi, Roberto}} \yr{2016}  \at{Microfluidic studies of biofilm formation in dynamic environments}.  \jt{J. Bacteriol.}  \bvol{198}~(19),  \pg{2589--2595}.

\bibitem[Yoest {\em et~al.\/}(2022)Yoest, Buggeln, Doan, Johnson, Berman, Mitchell \& Solomon]{Solomon_SwIMs_2022}
{\sc \au{Yoest, Helena}, \au{Buggeln, John}, \au{Doan, Minh}, \au{Johnson, Payton}, \au{Berman, Simon~A.}, \au{Mitchell, Kevin~A.} \& \au{Solomon, Thomas~H.}} \yr{2022}  \at{Barriers impeding active mixing of swimming microbes in a hyperbolic flow}.  \jt{Front. Phys.}  \bvol{10}.

\bibitem[Zaferani {\em et~al.\/}(2018)Zaferani, Cheong \& Abbaspourrad]{Zaferani_PNAS_2018}
{\sc \au{Zaferani, Meisam}, \au{Cheong, Soon~Hon} \& \au{Abbaspourrad, Alireza}} \yr{2018}  \at{Rheotaxis-based separation of sperm with progressive motility using a microfluidic corral system}.  \jt{Proc. Natl. Acad. Sci. U.S.A.}  \bvol{115}~(33),  \pg{8272--8277}.

\bibitem[Zhan {\em et~al.\/}(2014)Zhan, Sardina, Lushi \& Brandt]{Zhan_JFM_2014}
{\sc \au{Zhan, Caijuan}, \au{Sardina, Gaetano}, \au{Lushi, Enkeleida} \& \au{Brandt, Luca}} \yr{2014}  \at{Accumulation of motile elongated micro-organisms in turbulence}.  \jt{J. Fluid Mech.}  \bvol{739},  \pg{22–36}.

\bibitem[Zimmer \& Riffell(2011)]{Riffell_PNAS_2011}
{\sc \au{Zimmer, Richard~K.} \& \au{Riffell, Jeffrey~A.}} \yr{2011}  \at{Sperm chemotaxis, fluid shear, and the evolution of sexual reproduction}.  \jt{Proc. Natl. Acad. Sci. U.S.A.}  \bvol{108}~(32),  \pg{13200--13205}.

\end{thebibliography}

\end{document}